\documentclass[%
 reprint,
superscriptaddress,
 amsmath,amssymb,
 aps,
 prl,
]{revtex4-2}

\usepackage{graphicx}
\usepackage{dcolumn}
\usepackage{bm}

\begin{document}

\preprint{APS/123-QED}

\title{ Universality in spectral condensation}

\author{Induja Pavithran}
 \affiliation{Department of Physics, IIT Madras, Chennai-600036, India}
\author{Vishnu R. Unni}%
\affiliation{Department of Mechanical and Aerospace Engineering, University of California San Diego, California 92093, USA}
\author{Alan J. Varghese}
\affiliation{Department of Aerospace Engineering, IIT Madras, Chennai-600036, India}
\author{D. Premraj}
\affiliation{Department of Aerospace Engineering, IIT Madras, Chennai-600036, India}
\author{R. I. Sujith}
\email{sujith@iitm.ac.in}
\affiliation{Department of Aerospace Engineering, IIT Madras, Chennai-600036, India} 
\author{C. Vijayan}
\affiliation{Department of Physics, IIT Madras, Chennai-600036, India}

\author{Abhishek Saha}
\affiliation{Department of Mechanical and Aerospace Engineering, University of California San Diego, California 92093, USA}
\author{Norbert Marwan}
\affiliation{Potsdam Institute for Climate Impact Research, Germany}
\author{J\"urgen Kurths}
\altaffiliation[Also at ]{Department of Physics, Humboldt University, Germany}
\altaffiliation[Also at ]{Institute for Complex Systems and Mathematical Biology, University of Aberdeen, United Kingdom}
\affiliation{Potsdam Institute for Climate Impact Research, Germany}



\date{\today}

\begin{abstract}

Self-organization is the spontaneous formation of spatial, temporal, or spatiotemporal patterns in complex systems far from equilibrium. During such self-organization, energy distributed in a broadband of frequencies gets condensed into a dominant mode, analogous to a condensation phenomena. We call this phenomenon spectral condensation and study its occurrence in fluid mechanical, optical and electronic systems. We define a set of spectral measures to quantify this condensation spanning several dynamical systems. Further, we uncover an inverse power law behaviour of spectral measures with the power corresponding to the dominant peak in the power spectrum in all the aforementioned systems. 

\end{abstract}

\maketitle



During self-organization, an ordered pattern emerges from an initially disordered state. In dynamical systems, a pattern can be any regularly repeating arrangements in space, time or both \cite{cross1993pattern}. For example, a laser emits random wave tracks like a lamp until the critical pump power, above which the laser emits light as a single coherent wave track with high-intensity \cite{hermann1978synergetics}. A macroscopic change is observed in the laser system as a long-range pattern emerges in time. Another example is the Rayleigh-B\'enard system. For lower temperature gradients, the fluid parcels move randomly. As the temperature gradient is increased, a rolling motion sets in and the fluid parcels behave coherently to form spatially extended patterns. The initial random pattern can be regarded as a superposition of a variety of oscillatory modes and eventually some oscillatory modes dominate, resulting in the emergence of a spatio-temporal pattern \cite{croquette1989convective, kelso1995dynamic}. 
Self-organization often results in the redistribution of energy from a wide range of frequencies to a few dominant modes to form periodic patterns. Such a condensation in the spectrum is analogous to the condensation phenomenon observed in classical and quantum systems and we call this phenomenon spectral condensation.

Bose-Einstein condensation (BEC) occurring in quantum systems is characterized by occupation of the same energy level by a large fraction of the particles as temperature approaches absolute zero \cite{davis1995bose, ketterle1999experimentBEC}.
The transition to the condensate state, where the particles act collectively as a wave, can be viewed as the emergence of an ordered pattern from a disordered state of particles having different energy. 
Researchers have reported the observation of light condensation with the emission spectrum collapsing to the frequency of the lowest-loss mode \cite{fischer2012does, 
klaers2010bose, conti2008condensation}.
Similarly, by drawing parallels to BEC, condensation phenomenon has been used to explain several dynamical transitions where an ordered final state is achieved from an initially disordered state even in classical systems \cite{sun2012observation}. For instance, a population of coupled oscillators forms a dynamical condensate where the condensation phenomenon leads to global synchronization among the group of oscillators \cite{zanette1998condensation}. Likewise, the framework of BEC has been utilized in predicting the competitive dynamics in the evolution of complex networks \cite{bianconi2001bose}.

During self-organization driven by positive feedback leading to an ordered state in fluid mechanical, optical or electronic systems, we observe spectral condensation in the power spectrum of the appropriate system variables (in the emission spectrum for the optical system). While pattern formation in such systems has been studied extensively, universal characteristics of pattern formation or condensation in these systems garnered less attention. In conditions where the system is influenced by external noise or inherent fluctuations, the emergence of such a periodic pattern can be gradual as the parameter is varied. 
In this study, we quantify spectral condensation across various systems by defining a set of spectral measures based on the power spectrum.
The power of the dominant mode is found to scale with these spectral measures following an inverse power law. From experimental observations, we find that systems exhibiting self-organization driven by positive feedback follow unique way of spectral condensation in spite of different underlying physical mechanisms.

\begin{figure*}
\begin{center}
\includegraphics[scale = 0.30] {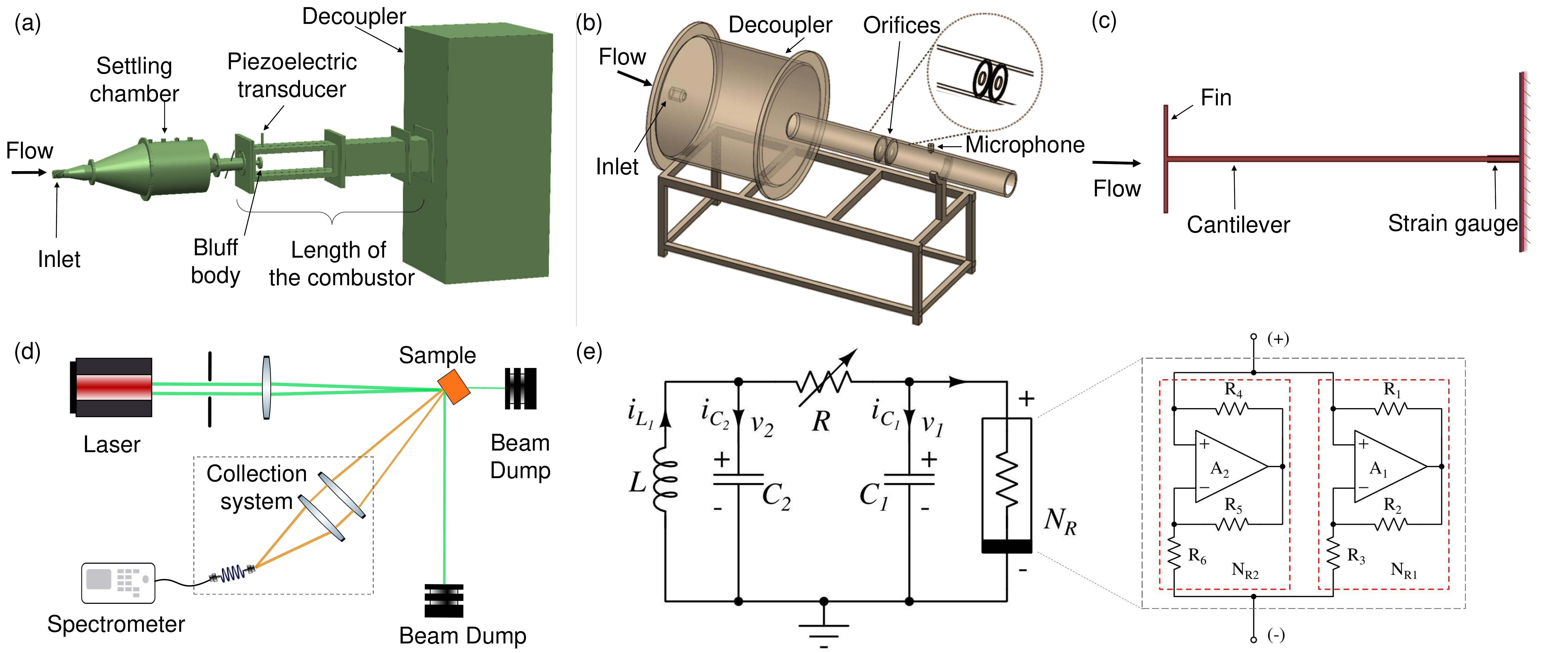}
\setlength{\belowcaptionskip}{-20pt}
\caption{Schematic of the experimental setups. (\textbf{a-c}) Three fluid mechanical systems exhibiting oscillatory instabilities, namely, a thermoacoustic, an aeroacoustic and an aeroelastic system respectively. In these fluid mechanical systems, the Reynolds number ($Re$) is the control parameter. (\textbf{d}) An optical system for random lasing, wherein we observe a transition towards narrow-band lasing like emission as we increase the excitation pulse energy (EPE) of the laser source. (\textbf{e}) Electronic circuit showing Chua’s circuit along with a zoomed detailed view of the Chua’s diode. The variable resistor ($R$) is varied to obtain the transition from a fixed point to limit cycle oscillations. In all these systems, we acquire data for different values of the respective control parameters. We measure the acoustic pressure fluctuations for thermoacoustic and aeroacoustic systems while the strain on the cantilever beam is acquired for the aeroelastic system. For the random laser, the output emission is collected using a fibre optic spectrometer and the voltage ($v_1$) is measured for the electronic circuit. The dimensions of different experimental setups are not to scale.}
\label{fig1}
\end{center}
\end{figure*}

Fluid mechanical systems examined in this study include thermoacoustic, aeroacoustic and aeroelastic systems which exhibit transition to oscillatory instabilities upon varying a control parameter. In a thermoacoustic system, the positive feedback between the acoustic field and the reactive flow field inside a confinement can cause an emergence of coherent dynamics in the flow field, which manifests as large amplitude self-sustained oscillations in pressure and heat release rate. This dynamical state is known as thermoacoustic instability \cite{juniper2018sensitivity}. The large amplitude oscillations encountered are detrimental to the structural integrity of practical systems such as rockets and gas turbine engines\cite{lieuwen2005combustion, fleming1998turbine}. 
Aeroacoustic instability is another oscillatory instability which arises due to the interaction between the acoustic field and the vortices in a turbulent flow \cite{flandro2003aeroacoustic}. Aeroelastic instability occurs as a result of the coupling between the turbulent flow and the structural elements of the system \cite{hansen2007aeroelastic}. Further details about the experiments are provided in Supplementary material. We analyze the sharpening of the dominant peak in the power spectrum of a fluctuating system variable in the following cases: a thermoacoustic system with different flame holding mechanisms and different combustor lengths (Fig.~\ref{fig1}a), an aeroacoustic system (Fig.~\ref{fig1}b) and an aeroelastic system (Fig.~\ref{fig1}c). Thermoacoustic system with different combustor lengths helps to achieve different characteristic time scales. Further, the two flame holding mechanisms in the combustor causes different mechanisms of thermoacoustic instability.

\begin{figure}
\begin{center}
\includegraphics[scale = 0.68] {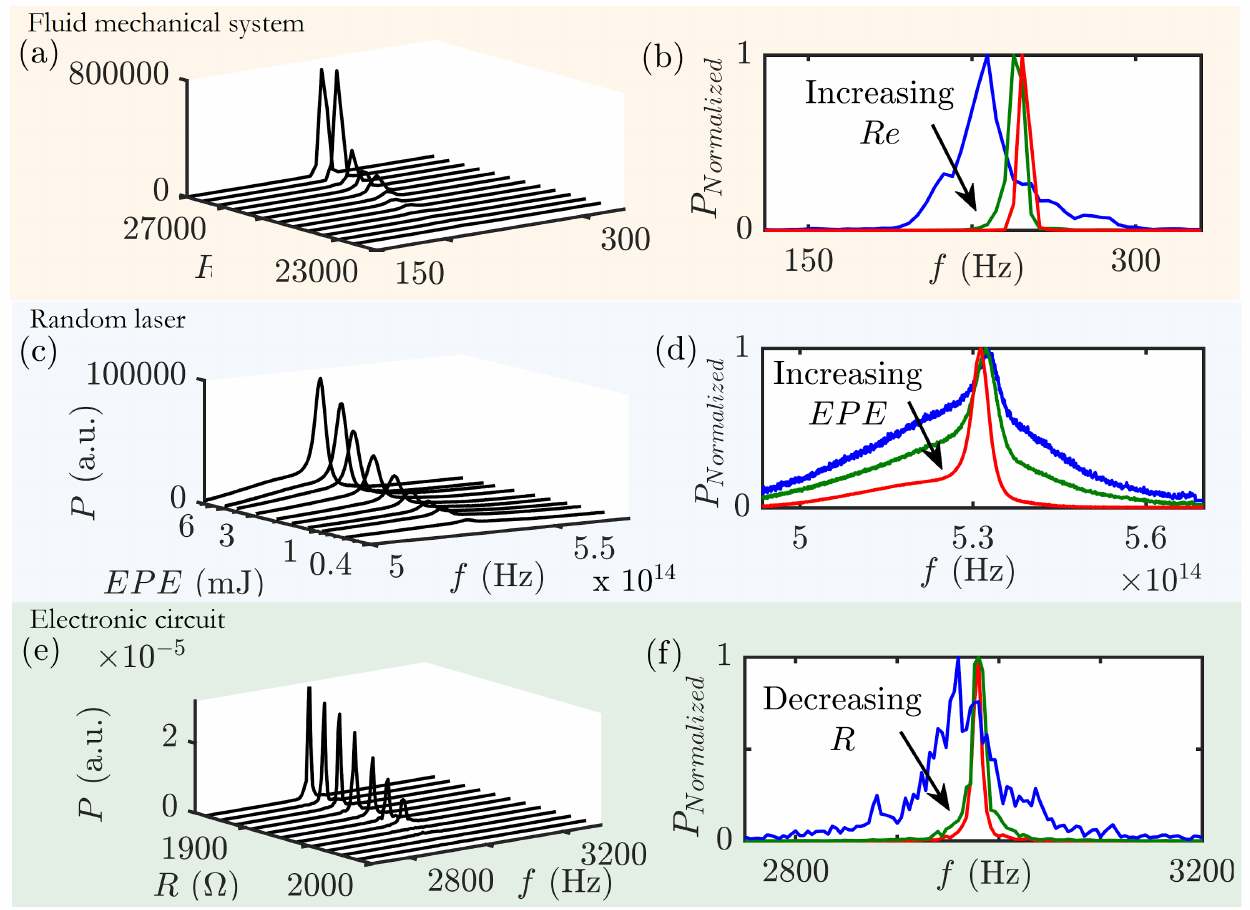}
\setlength{\belowcaptionskip}{-20pt}
\caption{Evolution of the spectrum for fluid mechanical, optical and electronic systems. The evolution of power spectrum with variation in the corresponding control parameter is presented on the left side and the normalized spectra for each system is given on the right side of the panel. (\textbf{a, b}) Power spectra obtained using Fast Fourier Transform (FFT) of the acoustic pressure fluctuations for a laboratory-scale bluff body stabilized combustor of length 700 mm. The power spectra exhibit an increasing dominance of a single peak on approaching oscillatory instabilities (for increasing $Re$) in all the fluid mechanical systems discussed in this paper, and hence this figure is a representative example. (\textbf{c, d}) Emission spectra of the random laser as we progressively increase EPE. The power spectrum is obtained by multiplying the photon count per second for each wavelength with its respective energy. The broad spectrum starts to become a narrow lasing-like peak with increase in EPE. (\textbf{e, f}) Power spectra obtained using FFT of the voltage signal ($v_1$) measured from Chua’s circuit and the corresponding normalized spectra are shown respectively. The peak sharpens during the transition to limit cycle oscillations as the resistance ($R$) is decreased. 
The power spectra using FFT are plotted for a resolution of 4 Hz for visualization purpose.}
\label{fig2}
\end{center}
\end{figure}

All the three fluid mechanical systems exhibit a transition to oscillatory instability as we increase $Re$. 
Here, we present the evolution of the power spectrum only for a representative set of data (Fig.~\ref{fig2}a), although all aforementioned cases of the fluid mechanical systems have been analyzed. 
The power spectrum has a broad peak for low values of $Re$. We observe the transition of power spectra from a broad peak to a sharp one as we approach the onset of the oscillatory instability. Each spectrum is normalized with its maximum amplitude to emphasize the narrowing of the peak (Fig.~\ref{fig2}b).

The optical system chosen for this study is a random lasing system (Fig.~\ref{fig1}d).
Unlike conventional lasers, the lasing action in random lasers is achieved by strong multiple scattering in the optical gain medium. The large number of scatterers which are dispersed in the gain medium causes the light rays to scatter multiple times before they exit the gain medium \cite{cao2003lasing, gummaluri2018stokes}.
The emission spectrum of a random laser, upon excitation by a pulse of suitable wavelength, is acquired using a fibre optic spectrometer. 
There is an appreciable narrowing in the emission profile (Fig.~\ref{fig2}c) with the increment in  excitation pulse energy (EPE), as is evident in Fig.~\ref{fig2}d where each spectrum is normalized with its maximum power. 

To study spectral condensation in electronic systems, we select Chua’s circuit (Fig.~\ref{fig1}e) which has become a paradigm for chaos \cite{chua1987linear, lakshmanan1996chaos}. It consists of two capacitors, an inductor, a resistor and one nonlinear element known as Chua’s diode.
The system exhibits period-doubling bifurcation from a fixed point to chaos with change in $R$. Here, we focus on the transition from a fixed point to a period-1 limit cycle. In experiments, external noise or inherent fluctuations including thermal fluctuations of the electronic devices, their inaccuracies and electromagnetic interference will make the fixed point noisy \cite{prebianca2018describing}. Thus, for the conditions for which a fixed point is expected, we observe low amplitude noisy oscillations with a broad peak in the power spectrum centered around the natural frequency.
During this transition to limit cycle (noisy Hopf bifurcation), we find a narrowing of the power spectrum (Fig.~\ref{fig2}e,~f) akin to that observed in fluid mechanical and optical systems. 

Next, we quantify the sharpening of the power spectrum during spectral condensation by defining `spectral measures'. The general expression for the spectral measure is:
\begin{eqnarray}
M_{m, n}^{x, y}=\left[\int\displaylimits_{-\delta F}^{+\delta F} \frac{P(F)}{P_0}\left|\frac{F}{f_{0}}\right|^{m} d F\right]^{x} \nonumber \left[\int\displaylimits_{-\delta F}^{+\delta F} \frac{P(F)}{P_0}\left|\frac{F}{f_{0}}\right|^{n} d F\right]^{y}.
\end{eqnarray}

Here, $P(F)$ represents the power corresponding to the modified frequency $F = f - f_0$, where $f$ is a variable indicating the frequency of oscillations, and $f_0$ is the frequency corresponding to the dominant peak in the power spectrum. $P_0 = P(f_0)$ and the indices $m,\ n,\ x\ \& \ y$ of the spectral measure are chosen to be positive integers. As our interest is to study the condensation towards a single peak, we compute the spectral measures $M_{m, n}^{x, y}$ only in the neighbourhood of width $\delta F$ centered at $f_0$. We set $\delta F$ to $f_{0} / 5$, based on our analysis of a collection of data with vastly different values of $f_0$. Also, the amplitude of the peak reduces significantly within this range.
Variations in the choice of $\delta F$ can be tried out, based on the appearance of the spectrum, such that it covers the spread of the peak during condensation. According to the definition of the spectral measures, $M_{m, n}^{x, y}$ decreases as the peak gets sharper. 
In this study, we present the analysis of three representative spectral measures, $M_{2,0}^{1,0},\ M_{2,0}^{1,1}$ $\&$ $M_{4,4}^{1,1}$. Note that $M_{2,0}^{1,0}$ can be considered as the second moment of the power spectrum in the $\delta F$ neighbourhood of $f_0$, whereas, $M_{2,0}^{1,1}\ \& \ M_{4,4}^{1,1}$ are the products of higher moments of the distribution. Higher moments give more weightage to the tail ends of the spectrum and thus its variation indicates how the broad tails diminish.

\begin{figure*}
\begin{center}
\includegraphics[scale = 0.55] {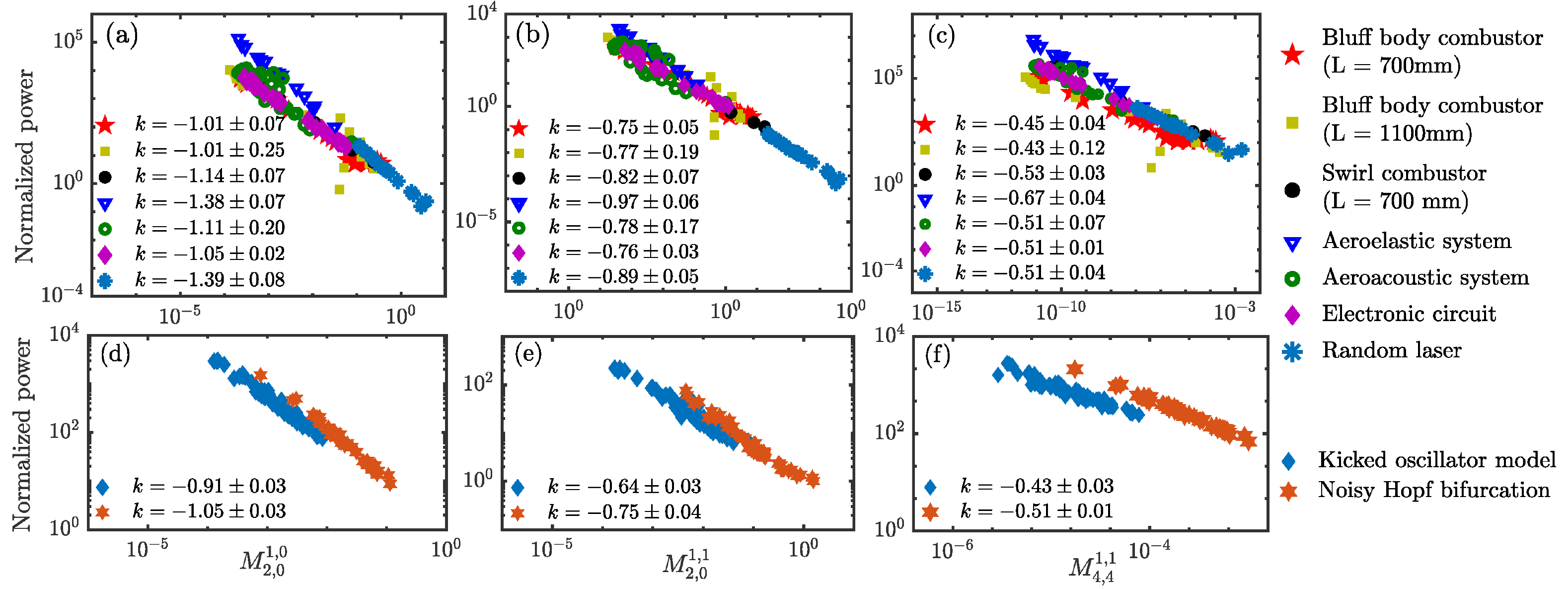}
\setlength{\belowcaptionskip}{-20pt}
\caption{The scaling behaviour of spectral measures with the peak power during spectral condensation in experimental systems and models. Variation of the normalized power ($P_0/P_N$) corresponding to the dominant mode with the representative spectral measures ($M_{2,0}^{1,0},\ M_{2,0}^{1,1}$ $\&$ $M_{4,4}^{1,1}$) plotted in double logarithmic scale  \textbf{(a-c)} for the data acquired from experiments conducted in different systems and (\textbf{d-f}) for the data generated from two models. The extent of spectral condensation and the peak power differs by orders of magnitude across these systems. Hence, we rescale the power corresponding to the dominant peak ($P_0$) as $P_0/P_N$ to show the lines, $\log (P_0) = k\ \log \left(M_{m, n}^{x, y}\right) + C$, in the same plot. The normalization factor, $P_N$, is the estimated value of peak power for $M_{m, n}^{x, y} = 1$ obtained by extrapolating the line  $\log (P_0) = k\ \log \left(M_{m, n}^{x, y}\right) + C$ for each system. This choice of $P_N$ forces all lines to have $C=0$ for $M_{m, n}^{x, y} = 1$. We observe an inverse power law behaviour for all the spectral measures in the experiments as well as in the models. The uncertainties in the power law exponent are shown for $95\%$ confidence intervals.} 
\label{fig3}
\end{center}
\end{figure*}

We uncover an inverse power law relation between the spectral measures and the power corresponding to the dominant peak (Fig.~\ref{fig3}a-c) during spectral condensation. All the data sets for the fluid mechanical, the optical and the electronic systems collapse to an inverse power law scaling in spite of the different physics involved in the process of condensation.
We also present the analysis of data obtained from two models: kicked oscillator model \cite{seshadri2016reduced} and noisy Hopf bifurcation model \cite{noiray2017linear} (detailed descriptions are given in Supplementary material). Both the models exhibit a transition from low amplitude aperiodic oscillations to a high amplitude limit cycle, thereby a condensation behaviour is observed in the power spectrum. We detect a similar scaling relation between $M_{m, n}^{x, y}$ and $P_0/P_N$ (Fig.~\ref{fig3}d-f) as observed in experiments. 
This inverse power law behaviour appears to be a universal characteristic of spectral condensation and the experimentally observed value for the power law exponent ($k$) corresponding to the spectral measures $M_{2,0}^{1,0},\ M_{2,0}^{1,1}$ $\&$ $M_{4,4}^{1,1}$ are around -1.12 $\pm$ 0.13, -0.7 $\pm$ 0.08 and -0.50 $\pm$ 0.06 respectively (averaged across systems). 
The exponent ($k$) is found to reduce for the higher indices of the measure and $k$ for higher moments have much narrower dispersion across different systems (refer Supplementary material). 


The existence of multiple invariant exponents motivates us to think about the existence of a universal form for the power spectrum in the neighbourhood of $f_0$. The power law relations indicate that given a distribution of power over a range of frequencies, the spectral measures at all levels of spectral condensation is already determined by the inverse power law relations. 
Further, the power spectrum decays away from $f_0$ and this decay is steeper for a sharp peak with higher amplitude. Thus, we consider a functional form for the power spectrum which is a function of $F$ and has $P_0$ and $f_0$ as two parameters, and is as follows:
\begin{equation}
P(F) = P_{0} e^{\left[-\left(P_{0}\right)^{\alpha}\left(\frac{F}{f_{0}}\right)^{\beta}\right]},
\label{eq2}
\end{equation}
where all the symbols retain their definitions. Here, both $\alpha$ and $\beta$ have to be strictly positive.
By comparing with the experimentally obtained values of power law exponents ($k$) for a set of spectral measures (for combinations of $m,\ n,\ x\ \& \ y$), we estimate the optimal values of the parameters $\alpha$ and $\beta$ iteratively as $\alpha = 0.125 \pm 0.017$ and $\beta = 0.317 \pm 0.024$ respectively.


In summary, we define spectral measures to compare and quantify spectral condensation in different systems and we uncover a universal route through which spectral condensation occurs in fluid mechanical, optical and electronic systems. The dominant peak in the power spectrum sharpens with an increase in peak power following inverse power law relations with the spectral measures. Interestingly, the scaling exponents are found to be within a small range across all the systems studied. From a practical viewpoint, these spectral measures can be used as a system independent method to quantify dynamical transitions in systems where an emergent periodic behaviour is observed. During spectral condensation in fluid mechanical systems, the redistribution of energy to a dominant mode causes high amplitude periodic oscillations which can have catastrophic effects on the system. In such cases estimating the peak power during oscillatory instability will help to design control strategies to mitigate oscillatory instabilities \cite{pavithran2020universality}. 
In future studies, it will be interesting to study this scaling behaviour for biological systems. 

\begin{acknowledgments}
We acknowledge the Department of Science and Technology, Government of India for the funding under the grant Nos.: DST/SF/1(EC)/2006 (Swarnajayanti Fellowship) and JCB/2018/000034/SSC (JC Bose Fellowship). We thank Dr. V. Nair (IIT Bombay), Dr. G. Thampi (Cochin University of Science and Technology), Mr. A Misra (Weizmann Institute of Science), V. Godavarthi (UCLA) and our colleagues from the IIT Madras for critical reading of the manuscript and useful discussions, Mr. Manikandan, Mr. Midhun, Mr. Thilagraj and Mr. Anand for the technical assistance during the experiments.
I.P. is grateful to Ministry of Human Resource Development, India and Indian Institute of Technology Madras for providing research assistantship.
\end{acknowledgments}



\providecommand{\noopsort}[1]{}\providecommand{\singleletter}[1]{#1}%

\end{document}